\documentstyle[11pt]{article}
\begin{document}     
\topskip 0.0  cm     
\def\cao{\c c\~ao\ }
\def\coes{\c c\~oes\ }
\def\ii{\'\i}
\def\CAO{\c C\~AO\ }
\def\COES{\c C\~OES\ }
\def\fpe{f\'\i sica de part\'\i culas elementares\ }
\def\nao{n\~ao\ }
\def\NAO{N\~AO\ }
\def\sao{s\~ao\ } 
\def\SAO{S\~AO\ } 
\def\ao{\~ao\ }
\def\AO{\~AO\ }
\def\be{\begin{equation}\ }
\def\ee{\end{equation}}
\def\lsim{\lower.7ex\hbox{$\;\stackrel{\textstyle<}{\sim}\;$}\ }
\def\gsim{\lower.7ex\hbox{$\;\stackrel{\textstyle>}{\sim}\;$}\ }

%\begin{flushright}
%\begin{tabular}{c c}
%& {\normalsize  IFT-NI.xxx/2000}\\
%& {\normalsize  xxx }\\
%& \today
%\end{tabular}
%\end{flushright}

%\vspace{2cm}
%\centerline{\Large\bf Motiva\c c\~ao do Ensino de F\'\i sica: }
%\vskip .2cm

\begin{center}
{\Large\bf  Mec\^anica Relacional: A prop\'osito de uma resenha 
%~\footnote{Vers\~ao aumentada das 
%Notas Internas IFT-NI.002/95. Submetida para publica\cao na Revista 
%Brasileira de Ensino de F\ii sica.} 
}
\\
%(Relational Mechanics: xxx )
\end{center}

\vskip 2 cm
\centerline{ {\bf C. O. Escobar $^a$ e V. Pleitez $^b$} 
%\footnote{E-mail:vicente@ift.unesp.br} 
}
\vskip 1.0cm
\centerline{ $^a$ Instituto de F\' {\i}sica Gleb Wataghin}
\centerline{Universidade Estadual de Campinas, UNICAMP}   
\centerline{13084-971 -- Campinas, SP, Brazil}

\vskip .5cm
%$^{(b)}$
\centerline{ $^b$ Instituto de F\'\i sica Te\'orica--
Universidade Estadual Paulista }
\centerline{Rua Pamplona, 145 }
\centerline{011405-900--S\~ao Paulo, SP, Brazil} 

\vskip 3cm
\centerline{\bf RESUMO}
Neste artigo fazemos uma an\'alise cr\'\i tica \`a proposta da Mec\^anica
Relacional tal como apresentada no livro de mesmo nome, objeto de uma resenha
recente nesta revista.

\vskip .5cm
\centerline{\bf ABSTRACT}
We present a critical analysis of what is called Relational Mechanics, as it
has been presented in a book thus entitled, which has 
been recently reviewed in this journal.

\newpage

\section{Introdu\c c\~ao}
\label{sec:intro}

Ainda que n\~ao seja parte do dia-a-dia de um pesquisador,
de vez em quando a quest\~ao do {\sl m\'etodo cient\'\i fico} aparece
para ser considerada, mesmo que seja de maneira breve, 
instant\^anea. Afinal, tantas coisas para fazer e a vida \'e t\~ao curta! 
No entanto,  seja motivado pela leitura de um trabalho ex\'otico colocado 
na rede eletr\^onica de {\it preprints}, seja pelo artigo confuso de 
uma revista especializada, vez ou outra somos levados a nos perguntar: 
O que distingue a ci\^encia de outras atividades? Como fazem os cientistas
para e\-li\-mi\-nar ou confirmar as teorias? 
\'E poss\'\i vel distinguir {\it ci\^encia}
da {\it pseudoci\^encia}? Existe {\it ci\^encia patol\'ogica}? Tem alguma
import\^ancia estas quest\~{o}es?
Por ex\-em\-plo, este tipo de preocupa\c{c}\~{a}o 
teria alguma implica\c c\~ao na nossa vida de 
pesquisador? \'E (ou deve ser) a ci\^encia conservadora?
\'E freq\"uente lembrar dos casos de persegui\c c\~ao cient\'\i fica:
Giordano Bruno, Galileu ou, de pelo menos cegueira coletiva da comunidade
cient\'\i fica: Boltzmann por exemplo ou, mais recentemente, Alfred Wegener
~\cite{allegre}. Deveria isso imobilizar a comunidade
cient\'\i fica? afinal quem \'e essa comunidade?~\footnote{Estes casos
n\~ao s\~ao exatamente como os manuais descrevem mas n\~ao \'e nosso 
objetivo aqui dar detalhes deles.}

A maioria das atividades que podemos classificar com os adjetivos adicionais
ao substantivo {\it ci\^encia}, mencionados no par\'agrafo anterior,
s\~ao re\-a\-li\-za\-das fora das universidades. 
Assim, com exce\c c\~ao de alguns cientistas como 
C. Sagan~\cite{sagan}, os pesquisadores n\~ao 
se d\~ao o trabalho de discutir e criticar essas atividades
da mesma maneira como criticam os pr\'oprios trabalhos cient\'\i ficos. 
Afinal, uma das caracter\'\i sticas
do dia-a-dia da ci\^encia \'e essa tens\~ao entre propostas
alternativas como explica\c c\~ao dos fen\^omenos naturais.
Mas, e quando isso acontecer 
numa universidade? S\~ao as cr\'\i ticas necess\'arias? Violariam a 
liberdade acad\^emica? A liberdade acad\^emica deve ser ampla e irrestrita?
Se sim, \'e isso compat\'\i vel com um bom crit\'erio de  
utiliza\c c\~ao dos fundos p\'ublicos? 

Recentemente foi publicado o livro {\sl Mec\^anica Relacional} (MR)~\cite{mr}.
Nesse livro pretende-se colocar uma nova vis\~ao da mec\^anica. Seria
mais um livro de ensino dessa disciplina ou um livro de divulga\c c\~ao
cient\'\i fica? Nenhum desses casos, sen\~{a}o 
vejamos. Um livro que afirme no pref\'acio: 

\begin{quotation}
Este livro tem como objetivo
apresentar as propriedades e carater\'\i sticas desta {\bf nova vis\~ao da 
mec\^anica} [$\cdots$] fica f\'acil fazer uma compara\c c\~ao com as 
{\bf vis\~oes anteriores}~\footnote{ Os negritos s\~ao nossos.} 
do mundo ({\bf newtoniana e einsteiniana}),~\footnote{Observe-se o tempo
passado com rela\c c\~ao a f\'\i sica newtoniana e einsteiniana.}
\end{quotation}
e que, al\'em disso, \'e editado 
pelo Centro de L\'ogica, Epistemologia e Hist\'oria da Ci\^encia da UNICAMP
n\~ao pode passar desapercebido pela comunidade cient\'\i fica do pa\'\i s. 
Ele deve ser analisado, comentado, criticado pelos cientistas da 
mesma forma que o s\~ao as teorias e resultados experimentais
da ci\^encia {\it normal}. 
N\~ao \'e poss\'\i vel que algu\'em chegue dizendo que 
as vis\~oes de Newton e Einstein est\~ao erradas e ningu\'em da comunidade
dos f\'\i sicos diga nada. Confirme-se e aceite-se seu impacto
na f\'\i sica e ci\^encias afins ou coloque-se esta obra em merecido 
ostracismo.
Esperamos deixar claro neste artigo que, corretamente analisado,
o assunto levantado pelo livro em quest\~ao sequer pol\^emico \'e. 
Por\'em, depois das 
caracter\'\i sticas acima mencionadas, o livro tem de ser analisado 
criticamente. Tamb\'em, acrescente-se, de maneira definitiva.

\'E necess\'ario saber se de fato representa uma vis\~ao nova da f\'\i sica
porque, se for verdade, j\'a imaginaram, leitores? Ter\'\i amos que rever 
tudo que foi feito nas \'ultimas d\'ecadas, v\'arios pr\^emios Nobel 
deviam ser devolvidos, o Brasil estaria na vanguarda da ci\^encia.
Mas, e se n\~ao fosse? Seria um exemplo de {\it ci\^encia patol\'ogica}? 
Enfim ... definitivamente n\~{a}o pode passar sem ser percebido, ainda que
n\~{a}o seja pol\^emico.   

J\'a foi feita uma resenha sobre o referido livro, publicada nesta 
re\-vis\-ta~\cite{soares}, da{\'\i} a  ``resenha'' do t\'\i tulo. 
Nessa resenha n\~ao se poupam elogios \`a nova vis\~ao da f\'\i sica 
pretendida no livro MR. No entanto,
\'e interessante notar que na vers\~ao publicada dessa resenha 
foi acrescentada uma nota de rodap\'e onde se agradece a um \'arbitro 
an\^onimo, o qual pedia para o autor da resenha ler o livro de 
A. Pais~\cite{ap1}. Nesse livro encontra-se uma hist\'oria mais 
detalhada sobre a influ\^encia do princ\'\i pio de Mach no pensamento 
de Einstein. No entanto, a leitura do livro de Pais deveria ter induzido 
o autor da resenha a rev\^e-la toda e mesmo mudar sua opini\~ao sobre 
o livro. Mas no pr\'oprio livro MR, apenas s\~ao citadas as 
palavras de Einstein sobre a influ\^encia que Mach teve sobre ele num certo
per\'\i odo de sua vida.
Omite-se outras, e que n\'os incluimos aqui, nas quais Einstein rev\^e a sua 
posi\c c\~ao com rela\c c\~ao ao {\it princ\'\i pio de 
Mach}.~\footnote{Dever\'\i amos dizer, em geral, da {\it filosofia} de Mach.}

Assim, \'e nosso objetivo fazer uma cr\'\i tica 
\`a proposta da MR baseada nas teorias
cient\'\i ficas desenvolvidas nos \'ultimos 100 anos, mais ou menos. 
Tentamos deixar claro para o leitor que: 1) n\~ao
\'e verdade que as teorias da relatividade especial e geral (TRE e TRG, 
respectivamente) estejam erradas, elas s\~ao de fato muito bem 
verificadas experimentalmente; al\'em do que, 
conceitualmente elas t\^em permitido avan\c cos t\'ecnicos e te\'oricos 
em diversas \'areas como a astronomia, a astrof\'\i sica e, principalmente,
na \'area da f\'\i sica das intera\c c\~oes fundamentais. 2) \'E sim, a MR
que n\~ao descreve os fen\^omenos naturais observados. 

Nosso objetivo n\~ao \'e convencer o autor do livro que a sua proposta 
n\~ao concorda com os dados experimentais, mas procurar convencer o leitor, 
que por pouca familiaridade com as teorias da f\'\i sica do Sec. XX
pode pensar estar diante de uma proposta que na sua opini\~ao \'e, na pior 
das hip\'oteses, pelo menos ``cient\'\i fica'', perceber por si s\'o
que o que n\'os colocamos aqui \'e correto: 
n\~ao s\'o as cr\'\i ticas \`as TRE e TRG est\~ao erradas, mas a pr\'opria
MR est\'a h\'a muito tempo descartada pela experi\^encia.

Claro, n\~ao esperamos que apenas a leitura desta resenha seja suficiente 
para tal efeito. Ser\'a necess\'ario que o leitor que ainda tiver 
d\'uvidas procure consultar algumas das refer\^encias
aqui citadas que poder\~ao ser-lhe de utilidade, ainda que n\~ao pretendamos
ser exaustivos nesse aspecto.

Na Sec.~\ref{sec:mach} revisamos o princ\'\i pio de Mach visando esclarecer
qual foi a sua influ\^encia sobre Einstein. Ficar\'a claro que 
a partir de certo momento Einstein afastou-se dele. Verifica-se tamb\'em que
esse princ\'\i pio n\~ao seria
necess\'ario para a elabora\c c\~ao daquelas teorias (TRE e TRG). Na 
Sec.~\ref{sec:tr} enfatizamos que as teorias da relatividade, especial e 
geral, s\~ao teorias bem estabelecidas experimentalmente e que n\~ao
procedem as cr\'\i ticas a ambas feitas no  livro MR~\cite{mr}.  
Pelo contr\'ario, mostramos na Sec.~\ref{sec:mr} que a MR \'e a  teoria
que est\'a errada. Alguns coment\'arios finais est\~ao na Sec.~\ref{sec:con}. 

\section{O princ\'\i pio de  Mach}
\label{sec:mach}

Ernst Mach (1838-1916) foi um cientista polivalente, mas a sua maior 
influ\^encia foi na mec\^anica de fluidos e na filosofia. Foi um 
cr\'\i tico do conceito de espa\c co absoluto da mec\^anica newtoniana.
No pref\'acio da primeira edi\c c\~ao (alem\~a) do seu livro 
disse~\cite{mach}
\begin{quotation}
The present volume is not a treatise upon the application of the principles 
of mechanics. Its aim is to clear up ideas, expose the real significance of 
the matter, and get rid of metaphysical obscurities.
\end{quotation}

Essas palavras devem ser entendidas no contexto do empirismo radical
que muitos cientistas defendiam nas \'ultimas d\'ecadas do Sec. XIX.
A termodin\^amica era ent\~ao rainha absoluta como paradigma de ci\^encia.
Estava baseada apenas em quantidades que podiam ser medidas no laborat\'orio,
outro tipo de abordagem era considerado {\it metaf\'\i sico}. Isso
influenciaria muito o pensamento de Einstein, mas depois ele aceitou
que ``\'e a teoria que diz o que \'e observ\'avel e o que n\~ao 
\'e''~\cite{wh1}. 

A. Pais, referindo-se \`a critica que Mach fizera em seu livro de 
1883~\cite{mach} \`a mec\^anica de Newton, disse~\cite{ap2} 
\begin{quotation}
As mencionadas refer\^encias mostram que Mach reconhecia 
claramente os aspectos cl\'assicos da mec\^anica cl\'assica e que 
n\~ao esteve longe de exigir uma teoria da relatividade geral, 
isto h\'a cerca de meio s\'eculo antes!
\end{quotation}
Por\'em Mach disse em 1913~\footnote{Esta frase est\'a traduzida de maneira 
diferente por diferentes autores. Aqui queremos somente lembrar a 
intransig\^encia de Mach, sendo que ele mesmo acreditava ser 
``n\~ao dogm\'atico''.}  
\begin{quotation}
Devo $[\cdots]$ com igual intensidade
recusar ser precursor dos re\-la\-ti\-vis\-tas, como me retirei da 
cren\c ca atomista da atualidade.
\end{quotation}

A vis\~ao de Mach da mec\^anica est\'a bem resumida 
na afirma\c c\~ao de que 
\begin{quotation}
 quando $[\cdots]$ afirmamos que um corpo conserva sem altera\c c\~ao 
sua dire\c c\~ao e velocidade no {\it espa\c co}, nossa afirma\c c\~ao 
n\~ao \'e nem mais nem menos do que uma refer\^encia abreviada ao 
{\it universo} inteiro (os it\'alicos s\~ao de Mach)
$[\cdots]$  
\end{quotation}
Comentando as palavras de Mach acima Pais disse~\cite{ap3}:
\begin{quotation}
N\~ao encontramos no livro de Mach {\it como} se manifesta 
esta import\^ancia de todos os corpos, pois ele nunca prop\^os um esquema 
din\^amico expl\'\i cito para esta nova interpreta\c c\~ao da lei de 
in\'ercia.
\end{quotation}

Isso ainda \'e verdade: o princ\'\i pio de Mach n\~ao foi 
im\-ple\-men\-ta\-do de maneira consistente por nenhuma teoria.
O conhecido astron\^omo H. Bondi \'e bem claro a respeito~\cite{bondi}
\begin{quotation}
... the postulate of relativity of inertia (Mach's principle) is 
intelectually agreable in many ways, and seems to some authors to be 
inescapably true. Others regard it with suspicion, since it has
not been possible so far to express it in mathematical form (not even in
the general relativity), and since it has not so far been verified 
experimentally [$\cdots$] Even if Mach's principle is correct, other 
theories are therefore required to deal with experimental and observational 
invariance.
\end{quotation}

Como dissemos antes, a constru\c c\~ao da TRE foi muito influenciada pela 
filosofia pragm\'atica de Mach: foram usadas apenas quantidades 
pass\'\i veis de serem medidas. Einstein posteriormente tamb\'em se 
afastou dessa filosofia, mas n\~ao consideraremos isso aqui.
Por outro lado, o mesmo aconteceu com a TRG.
Em 1912, usando uma vers\~ao rudimentar da teoria da gravita\c c\~ao,
Einstein mostrou que se uma esfera oca massiva \'e acelerada  em torno de 
um eixo que passa pelo centro no qual se encontra uma massa inercial 
pontual, ent\~ao a massa inercial desta \'ultima \'e aumentada. 
Nas pr\'oprias palavras de Einstein~\cite{ap4}
\begin{quotation}
Esta [conclus\~ao] fornece plausabilidade \`a conjectura de que a in\'ercia 
{\it total} de um ponto com massa \'e um efeito que decorre da presen\c ca  
de todas as outras massas, gra\c cas a um tipo de intera\c c\~ao com estas 
\'ultimas [$\cdots$]. \'E este justamente o ponto de vista sustentado por Mach 
nas suas investiga\c c\~oes profundas sobre este tema.
\end{quotation}

Vemos que Einstein tinha de fato o princ\'\i pio de Mach como guia para
a constru\c c\~ao das teorias da  relatividade.~\footnote{De fato foi 
Einstein quem chamou a conjetura da origem da in\'ercia de Mach como o 
``Princ\'\i pio de Mach''.}  

Em 1917 Einstein, no que seria o primeiro trabalho da hist\'oria sobre 
cosmologia relativista~\cite{ae4}, ainda pensava de acordo com a 
cita\c c\~ao acima a respeito das id\'eias de Mach~\cite{ap4}. 
Ele ainda tentava implementar uma
origem inteiramente material da in\'ercia, isto \'e, que a m\'etrica 
$g_{\mu\nu}$ do espa\c co-tempo
seria determinada apenas pela mat\'eria~\cite{ap4}. De fato, nesse trabalho
Einstein introduz o ``termo cosmol\'ogico'' para estar em acordo com o 
princ\'\i pio de Mach, isto \'e, para ter um universo fechado, e tamb\'em para 
conseguir um universo homog\^eneo, isotr\'opico e est\'atico e tal que
$g_{\mu\nu}=0$ na aus\^encia de mat\'eria. 

Provavelmente a demostra\c c\~ao de de Sitter em 1917
sobre a exist\^encia de solu\c c\~oes das equa\c c\~oes da TRG 
no v\'acuo: $ g_{\mu\nu}\not=0$ e $T_{\mu\nu}=0$, isto \'e, solu\c c\~oes para
as equa\c c\~oes da TRG sem mat\'eria (que Einstein acreditava n\~ao 
ex\-is\-ti\-rem) \'e que come\c cou a minar sua  credibilidade nesse
princ\'\i pio. A outra motiva\c c\~ao, de um universo homog\^eneo, 
isotr\'opico {\it e} est\'atico, foi eliminada quando em 1922 A. Friedmann
demonstra que era poss\'\i vel um universo homog\^eneo e istr\'opico 
se ele estivesse se expandindo (e n\~ao est\'atico como assumia Einstein). 
Mas o acontecimento crucial foi ent\~ao a descoberta de de 
Sitter que as equa\c c\~oes de Einstein com o termo cosmol\'ogico tinham 
solu\c c\~ao mesmo no vazio: 
a in\'ercia \'e diferente de zero mesmo sem a presen\c ca 
da mat\'eria. Inicialmente Einstein, que antes tinha dito que ``um corpo num 
universo vazio n\~ao poderia ter in\'ercia'', objetou a solu\c c\~ao de 
de Sitter mas logo ele se convenceu que aquele tinha raz\~ao. N\~ao era
mais poss\'\i vel que $g_{\mu\nu}$ pudesse ser determinado completamente
pela mat\'eria.
Vemos ent\~ao que n\~ao se pode fazer uma cita\c c\~ao de Einstein de 1917 
sem levar em conta que alguns anos depois ele se convenceria de seu 
pr\'oprio erro!

Segundo Pais~\cite{ap7}
\begin{quotation}
Anos mais tarde, o entusiasmo de Einstein pelo princ\'\i pio de Mach
esmoreceu e, finalmente desapareceu.
\end{quotation} 

Por exemplo, em 1954 em uma carta a Felix Pirani ele disse~\cite{gh,ap5}
\begin{quotation}
Na minha opini\~ao nunca mais dever\'\i amos  falar do princ\'\i pio de 
Mach.
Houve uma \'epoca na qual pensava-se que os `corpos ponder\'aveis' eram 
a \'unica realidade f\'\i sica e que, numa teoria todos os elementos que 
n\~ao estiverem totalmente determinados por eles, deveriam ser 
escrupulosamente evitados. Sou consciente que durante um longo tempo 
tamb\'em fui influenciado por essa id\'eia fixa.
\end{quotation}

Pouco tempo depois ele disse~\cite{pas}
\begin{quotation}
...So, if one regards as possible, gravitational fields of arbitrary extension
which are not initially restricted by spatial limitations, the concept
`acceleration relative to space', then loses every meaning and with it
the principle of inertia together with the entire problem of Mach.
\end{quotation}

Em geral os cosm\'ologos aceitam o ponto de vista posterior de Einstein,
por exemplo, segundo Bondi~\cite{bondi} 
\begin{quotation}
for this reason
he introduced the so-called cosmological constant in the hope of
reconciling general relativity with Mach's principle [$\cdots$] This
hope was, however, not fulfilled.
\end{quotation}  

A origem da in\'ercia (das massas) continua a ser um ponto em aberto
em qualquer teoria fundamental das part\'\i culas elementares.
Assim segundo Pais~\cite{ap5}:
\begin{quotation}
Do meu ponto de vista, at\'e agora o princ\'\i pio de Mach n\~ao fez 
avan\c car decisivamente a f\'\i sica, e a origem da in\'ercia \'e, e 
continua a ser, o assunto {\it mais} obscuro na teoria de part\'\i culas e 
campos. O princ\'\i pio de  Mach pode, conseq\"uentemente  ter futuro,
mas n\~ao sem a teoria qu\^antica.
\end{quotation}

Podemos concluir que o princ\'\i pio de Mach n\~ao foi at\'e agora 
confirmado nem te\'orica nem experimentalmente. Que todas as teorias 
atuais n\~ao tenham sido capazes de implement\'a-lo pesa mais contra 
ele que contra as pr\'oprias teorias. 
O estudo da influ\^encia de Mach sobre Einstein pertence mais ao que Holton
chama de ``el peregrinaje filos\'ofico de Albert Einstein''~\cite{gh2}
\begin{quotation}
un peregrinaje desde una filosof\'\i a de la ciencia en la que el
sensacionalismo y el empirismo ocupaban una posici\'on central, hasta otra
que est\'a fundada en un realismo racional.
\end{quotation}

Definitivamente ent\~ao, a partir de um certo momento, Einstein e 
outros f\'\i sicos bem conhecidos n\~ao levaram mais em conta 
o princ\'\i pio de Mach como guia na constru\c c\~ao das suas teorias.

\section{As Teorias da Relatividade est\~ao erradas?}
\label{sec:tr}

N\~ao. Muito pelo contr\'ario. Vide, por exemplo, o amplo artigo de 
Will~\cite{will}, onde se resume os testes experimentais de ambas teorias 
da relatividade, a especial e a geral.
No caso da relatividade especial, que \'e sem d\'uvida a melhor testada,
Will diz~\cite{will2}:
\begin{quotation}
A lot of experiments in the high-energy laboratory have verified and 
reverified the validity of special relativity in the limit when gravitational 
effects can be ignored. Those experiments range from direct test of 
time-dilation to tests of esoteric predictions of Lorentz-invariant quantum 
field theory.
\end{quotation}
No entanto o autor de MR insiste~\cite{mr2}:
\begin{quotation}
Defendemos aqui que as teorias de Einstein n\~ao implementaram as id\'eias
de Mach e que a Mec\^ancia Relacional \'e uma teoria melhor do que
as de Einstein para descrever os fen\^omenos observados na 
natureza $[\cdots]$ Einstein e seus seguidores criaram muitos problemas 
com esta teoria.
\end{quotation}

A TRE tornou a hip\'otese do \'eter sup\'erflua, n\~ao mostrou que este 
n\~ao existia. Isso \'e t\'\i pico do conhecimento cient\'\i fico. 
O autor do texto MR n\~ao entendeu como funciona a ci\^encia.
A ci\^encia n\~ao mostra que os deuses da chuva e anjos carregando os 
planetas n\~ao existem. Ela apenas n\~ao usa essas hip\'oteses.
Claro, se acreditamos que existe uma realidade independente de n\'os mesmos 
e que \'e, pelo menos parcialmente, desvendada pela ci\^encia, ent\~ao 
o fato de o \'eter n\~ao ser necess\'ario para as teorias f\'\i sicas pode 
ser interpretado como indicativo de sua inexist\^encia.

As cr\'\i ticas do autor \`a TRE n\~ao s\~ao corretas e mostram a pouca 
fa\-mi\-lia\-ri\-da\-de dele com o tema. Por exemplo~\cite{mr4}% (mr, p.190)
\begin{quotation}
...h\'a muitos problemas com as teorias da relatividade especial e geral.
Enfatizamos alguns aqui.

1) elas s\~ao baseadas na formula\c c\~ao de Lorentz da eletrodin\^amica de 
Maxwell, formula\c c\~ao que apresenta diversas assimetrias como as apontadas 
por Einstein e muitos outros $[\cdots]$ H\'a uma teoria do eletromagnetismo 
que evita todos estas assimetrias de forma natural $[\cdots]$ a 
eletrodin\^amica de Weber...
\end{quotation}

\'E sabido, faz mais de 100 anos, que a eletrodin\^amica de Weber n\~ao 
\'e uma descri\c c\~ao dos fen\^omenos eletromagn\'eticos: 
na sua vers\~ao original n\~ao prev\^e
a exist\^encia de ondas eletromagn\'eticas! Da maneira como \'e comparada 
com ``a formula\c c\~ao de Lorentz da eletrodin\^amica de Maxwell'' parece 
que a de Weber \'e uma outra formula\c c\~ao da mesma. 
A formula\c c\~ao de Lorentz a que se refere o autor \'e a das equa\c c\~oes 
de Maxwell microsc\'opicas. Nela todos os fen\^omenos eletromagn\'eticos
podem ser vistos como sendo produzidos por portadores de cargas elementares
como os el\'etrons e os n\'ucleos at\^omicos. As equa\c c\~oes de Maxwell
macrosc\'opicas podem, em casos simples, ser deduzidas a partir das 
equa\c c\~oes de Maxwell-Lorentz. 
Na verdade \'e a formula\c c\~ao de Lorentz que
\'e generaliz\'avel para a mec\^anica qu\^antica relativista. 

A assimetria a que se refere o autor \'e aquela mencionada 
no primeiro artigo de Einstein de 1905 sobre a TRE~\cite{ae1,ae2}:
\begin{quotation}
 Como \'e sabido, a Eletrodin\^amica de Maxwell--tal como atualmente se 
concebe-- conduz, na sua aplica\c c\~ao a corpos em movimento,  
a {\bf assimetrias que n\~ao parecem ser inerentes aos 
fe\-n\^o\-me\-nos}.~\footnote{Os negritos s\~ao nossos.}
Consideremos, por exemplo, as a\c{c}\~{o}es eletrodin\^amicas entre um 
\'\i m\~{a} e um condutor. {\bf O fen\^omeno observ\'avel depende
unicamente do movimento relativo do condutor e do \'\i m\~{a}}, ao
passo que, segundo a concep\c c\~ao habitual, s\~ao nitidamente distintos
os casos em que o m\'ovel \'e um, ou outro, desses corpos. Assim, se for
m\'ovel o \'\i m\~{a} e o condutor estiver em repouso, estabelecer-se-\'a
em volta do \'\i m\~{a} campo el\'etrico com determinado conte\'udo 
energ\'etico, que dar\'a origem a uma corrente el\'etrica nas regi\~oes onde
estiverem colocadas por\c c\~oes do condutor. Mas, se \'e o \'\i m\~{a}
que est\'a em repouso e o condutor que est\'a em movimento, ent\~ao,
embora n\~ao se estabele\c ca em volta do \'\i m\~{a} nenhum campo el\'etrico,
h\'a no entanto uma for\c ca eletromotriz que n\~ao corresponde a nenhuma 
energia, mas que d\'a lugar a correntes el\'etricas
de grandeza e comportamento iguais \`as do primeiro caso,
produzidas por for\c cas el\'etricas--desde que, nos dois casos
considerados, haja identidade no movimento relativo.
\end{quotation}

Mais adiante, depois de apresentar a sua teoria, Einstein diz~\cite{ae3}
\begin{quotation}
Como se v\^e, na teoria que se desenvolveu, a for\c ca eletromotriz apenas 
desempenha o papel de conceito auxiliar, que deve a sua introdu\c c\~ao ao 
fato de as for\c cas el\'etricas e magn\'eticas n\~ao terem exist\^encia 
independente do estado de movimento do sistema de coordenadas.

\'E tamb\'em claro que a assimetria mencionada na introdu\c c\~ao, que surge 
quando se consideram as correntes el\'etricas provocadas pelo movimento 
relativo de um \'\i man e de um condutor, desaparece agora.
\end{quotation}

Vemos ent\~ao que o autor da MR n\~ao entendeu o argumento de Einstein no 
seu artigo de 1905. 
Hoje dir\'\i amos que as equa\c c\~oes de Maxwell, usando a 
nota\c c\~ao de 3-vetores, introduzida por Heaviside, n\~ao s\~ao 
manifestamente invariantes sob as transforma\c c\~oes de Lorentz. 
Mas essa assimetria n\~ao ocorre, como observado pelo pr\'oprio Einstein, 
nos fen\^omenos observados, como sabemos desde Faraday.
A assimetria desaparece porque no sistema de refer\^encia que acompanha
o condutor, do ponto de vista da TRE, temos tamb\'em um campo el\'etrico:
$\vec{E}'\propto v\times \vec{B}'$. A inter-rela\c c\~ao entre campos 
el\'etricos e magn\'eticos na eletrodin\^amica de Maxwell s\'o foi 
descoberta por Einstein. Ainda que a teoria microsc\'opica (que \'e de 
Lorentz mas continua sendo a eletrodin\^amica de Maxwell) seja 
relativisticamente invariante, o fen\^omeno mencionado foi percebido 
por Einstein mesmo. Assim, \'e apenas quando se descobre a invari\^ancia 
das equa\c c\~oes de Maxwell sob transforma\c c\~oes de Lorentz que a 
assimetria desaparece. Isso est\'a bem explicado em livros b\'asicos como 
o de Purcell~\cite{purcell} apenas para dar um exemplo.

Outro ponto que deve ser enfatizado \'e que o autor da MR 
n\~ao compreendeu a covari\^ancia geral, confundindo-a com a
covari\^ancia introduzida por Minkowski que se refere apenas \`as
transforma\c c\~oes de Lorentz~\cite{mr10}. %(mr, p.179). 
Na TGR sim, temos uma covari\^ancia 
geral, no sentido que as equa\c c\~oes s\~ao as mesmas em qualquer sistema 
de refer\^encia, inercial ou n\~ao. 

Como j\'a foi dito acima, a TRE n\~ao \'e verificada somente pelas 
experi\^encias diretas.
Todo o edif\'\i cio conceitual da f\'\i sica de part\'\i culas elementares
e as suas t\'ecnicas te\'oricas e experimentais est\~ao baseados nela.
Mesmo que algu\'em mostrasse que as experi\^encias cl\'assicas
n\~ao s\~ao suficientes para testar com a precis\~ao necess\'aria 
a teoria, esta n\~ao seria facilmente abandonada porque j\'a foi 
confirmada na pr\'atica em outras \'areas.

O mesmo ocorre com a TRG: 
nos anos de 1939-40 Einstein, com Leopold Infeld e Banesh Hoffmann, tratou
o problema do movimento de N corpos com a relatividade  geral. Segundo Misner 
et al.~\cite{gravitation}
\begin{quotation}
Equations $[\cdots]$ are called the Einstein-Infeld-Hoffman (EIH) equations
for the geometry and evolution of a many-body system. They are used widely
in analyses of planetary orbits in the solar system. For example, the 
Caltech Jet Propulsion Laboratory uses them, in modified form, to calculate
ephemerides for high-precision tracking of planets and 
spacecraft.
\end{quotation}
Vemos ent\~ao que j\'a existem aplica\c c\~oes da TRG (ver mais sobre isso
mais adiante).

Al\'em disso novos testes mais acad\^emicos s\~ao obtidos. 
Por exemplo, em 1993 R. A. Hulse e J. H. Taylor ganharam o pr\^emio Nobel de 
F\'\i sica:
``for the discovery of a new type of pulsar, a discovery that has opened up
new possibilities for the study of gravitation"~\cite{nobel93}.

Mas o que isso significa?
Bem, Hulse e Taylor observaram durante quase 20 anos, de 1975 a 1993,
um pulsar bin\'ario com o ex\'otico nome de PSR 
1913+16~\footnote{ PSR significa ``pulsar'' e 1913+16 especifica a 
posi\c c\~ao do pulsar no c\'eu.}--e que consiste de um par de 
estrelas de n\^eutrons, com um raio de algumas dezenas de quil\^ometros e
com massa da ordem da massa do Sol e com uma dist\^ancia relativa da ordem
da algumas vezes a dist\^ancia Terra-Lua 
girando ao redor de seu centro de massa. Eles determinaram que a perda 
de energia do sistema era consistente com os c\'alculos baseados na 
teoria da re\-la\-ti\-vi\-da\-de geral~\cite{rp}. 
Este foi um teste de TRG mais definitivo que
os tr\^es testes cl\'assicos: o perih\'elio de Merc\'urio, o desvio da luz 
pelo Sol e o atraso de rel\'ogios em campos gravitacionais. 
Estes testes estavam restritos
ao nosso sistema solar onde o campo gravitacional \'e fraco. 
Os resultados de Hulse e Taylor foram os primeiros 
testes de grande precis\~ao da TGR. Segundo a TRG, objetos em 
\'orbita, como no caso do pulsar acima mencionado, irradiam energia sob a 
forma de ondas gravitacionais (ondula\c c\~oes no espa\c co-tempo). Isto 
implica numa perda de energia do sistema que pode ser calculada u\-san\-do a 
TRG. 
Os resultados de Hulse e Taylor concordaram muito bem (\'e um n\'umero da 
ordem de $10^{-14}$ e medido com uma precis\~ao de 0.5\% !) 
com as previs\~oes te\'oricas da TGR.  

Em 100 anos de pr\^emios Nobel apenas em 6 ocasi\~oes n\~ao foi entregue.
Deste total, 27 est\~ao relacionados de alguma maneira com a relatividade
especial e pelo menos 1 com a TGR. Estes s\~ao:
P. A. M. Dirac (1933, teoria relativista do el\'etron), J. Chadwick (1935,
descoberta do n\^eutron), C. D. Anderson (1936, descoberta da anti-mat\'eria);
E. O. Lawrence (1939, inven\c c\~ao do ciclotron); W. Pauli (1945, 
princ\'\i pio de exclus\~ao); H. Yukawa (1949, pelo m\'esons $\pi$);
J. D. Cockcroft e E. T. S. Walton (1951, a\-ce\-le\-ra\-do\-res de 
part\'\i culas);
W. E. Lamb e P. Kush (1955, efeitos relativ\'\i sticos nos \'atomos);
C. N. Yang e T. D. Lee (1957, viola\c c\~ao da paridade); E. G. Segr\'e e
O. Chamberlein (1959, descoberta de anti-mat\'eria hadr\^onica: anti-pr\'oton);
E. P. Wigner (1963, princ\'\i pios de simetria); S-I. Tomonaga, J. Schwinger e
R. Feyman (1965, pela eletrodin\^amica relativista); H. A. Bethe (1967,
pelos mecanismo relativistas da cria\c c\~ao da energia nas estrelas);
L. W. Alvarez (1968, des\-co\-ber\-tas experimentais em f\'\i sica de 
part\'\i culas elementares); M. Gell-Mann (1969, contribui\c c\~oes 
te\'oricas \`a 
f\'\i sica de part\'\i culas elementares); B. Richter e S. C. C. Ting (1976, 
descoberta do quark $b$); S. Glashow, A. Salam e S. Weinberg (1979,
modelo de intera\c c\~oes eletrofracas); J. W. Cronin e V. L. Fitch (1980,
descoberta da viola\c c\~ao da simetria discreta CP); 
S. Chandrasekhar (1983, evolu\c c\~ao e estrutura das estrelas)
C. Rubbia e 
S. van der Meer (1984, descoberta dos b\'osons intermedi\'arios $W^\pm,Z^0$);
L. M. Lederman, M. Schwartz e J. Steinberg (1988, descoberta do segundo 
neutrino, $\nu_\mu$); G. Charpac (1992, detetores de part\'\i culas 
relativistas); M. Perl e F. Reines (1995, descobertas do l\'epton $\tau$ e 
dete\c c\~ao do neutrino do eletron, $\nu_e$, respectivamente);
G. 't Hooft e M. J. G. Veltman (1999, corre\c c\~oes qu\^anticas ao modelo 
eletrofraco de Glashow-Salam-Weinberg). Todas estas descobertas te\'oricas
ou experimentais somente t\^em sentido no contexto de teorias 
qu\^antico-relativistas.
Com rela\c c\~ao \`a TGR podemos colocar os j\'a acima mencionados 
R. A. Hulse e J. H. Taylor (1993, dete\c c\~ao indireta de ondas 
gravitacionais). 
N\~ao mencionamos aqui alguns resultados tamb\'em premiados que de maneira 
indireta usam a eletrodin\^amica de Maxwell,  
que sendo relativista poderia ser considerado como teste indireto da TRE.
Se as teorias da relatividade estivessem erradas todos esses
pr\^emios Nobel teriam que ser devolvidos. O leitor interessado 
pode visitar a p\'agina WWW da Funda\c c\~ao Nobel~\cite{nobelf}.

Com rela\c c\~ao \`a dilata\c c\~ao do tempo nas TRE e TRG,
o astr\^onomo real Martin Rees diz que mesmo n\~ao sendo percept\'\i vel
nos movimentos e tempos do dia-a-dia~\cite{rees} 
\begin{quotation}
Esse pequeno efeito [dilata\c c\~ao do tempo da TRE] foi agora, contudo, 
medido por 
experimentos com rel\'ogios at\^omicos com precis\~ao de um bilion\'esimo 
de segundo, e est\'a de acordo com as previs\~oes de Einstein... 
uma ``dilata\c c\~ao do tempo'' semelhante \'e causada pela gravidade: 
nas proximidades de uma grande massa, os rel\'ogios tendem a andar mais 
devagar....
essa dilata\c c\~ao deve ser levada em conta, juntamente com os efeitos
do movimento orbital, na programa\c c\~ao do notavelmente preciso sistema 
GPS (Global Positioning Satellite)...
\end{quotation}
De fato, atualmente o sistema GPS tem uma precis\~ao de milimetros, 
uma discord\^ancia de uma milhon\'esima de segundo implica num erro da
ordem de 300 metros!~\cite{herring}.

Al\'em disso as cada vez mais precisas medidas do fator $(g-2)_\mu$
s\~ao compat\'\i veis com dilata\c c\~oes da vida m\'edia do m\'uon de at\'e 
$\gamma=29.3$~\cite{g-2}
Isso mostra que o efeito nos m\'uons n\~ao apenas os observados na atmosfera
e o argumento do autor da MR n\~ao se sustenta (ver a pr\'oxima se\c c\~ao).  

Poder\'\i amos mencionar outras situa\c c\~oes onde fica claro o pouco 
conhecimento que o autor de MR tem das teorias da relatividade. 
Bastam mais um exemplos: o autor da MR n\~ao sabe que n\~ao existe 
``paradoxo dos g\^emeos'', n\~ao entendeu o atraso do 
rel\'ogio~\cite{mr12}. %.(mr, p.157)
N\~ao comentamos mais sobre este ponto porque \'e bastante bem considerado 
em livros elementares de relatividade~\cite{geroch}. 

\section{Est\'a a  Mec\^anica Relacional errada?}
\label{sec:mr}

Sim. Ap\'os criticar a TRG de Einstein por n\~ao ter implementado a rota de 
construir a teoria apenas em termos de dist\^ancias relativas, 
diz~\cite{mr3} 
\begin{quotation}
... como veremos neste livro, \'e poss\'\i vel seguir esta rota com sucesso 
utilizando uma lei de Weber para a gravita\c c\~ao.
\end{quotation}
 %(p.178)
Por que usar uma lei da gravita\c c\~ao baseada numa lei da eletrost\'atica
que n\~ao deu certo?  
Mesmo que algu\'em acredite na MR, dever-se-ia perguntar: por que essa 
for\c ca e n\~ao outra?  Assim, existiriam tantas MR quanto poss\'\i veis 
autores. O papel desempenhado pelas simetrias nas leis da
F\'\i sica n\~ao foi comprendido pelo autor da MR: ele ignora os
trabalhos de cientistas como E. Wigner, H. Weyl, C. N. Yang etc! 
As simetrias t\^em desempenhado um
papel importante na descoberta de novas leis da natureza, mas na MR
lemos que tudo isso n\~ao \'e necess\'ario na {\bf nova f\'\i sica} 
ali proposta! No momento que se abre m\~ao dos princ\'\i pios de simetria
tudo \'e v\'alido! 

A MR est\'a baseada em tr\^es postulados. Os dois primeiros s\~ao 
compat\'\i veis com as leis de Newton. J\'a o terceiro postulado, 
diz~\cite{mr5}
\begin{quotation}
A soma de todas as for\c cas de qualquer natureza (gravitacional, 
el\'etrica, magn\'etica, el\'astica, nuclear,...) agindo sobre qualquer 
corpo \'e sempre nula em todos os sistemas de refer\^encia.
\end{quotation}
Bom, sabemos que cada uma  das for\c cas mencionadas no postulado III
tem uma intensidade carater\'\i stica bem diferente. Por exemplo, a for\c ca 
gravitacional \'e $10^{-40}$ vezes mais fraca que a 
for\c ca eletromagn\'etica. Assim, se a sua soma se anula,
ent\~ao devem existir outras for\c cas tais que fa\c cam a soma ser zero.
Onde est\~ao essas for\c cas?

Compare com o postulado de Einstein ``a luz, no espa\c co vazio, se propaga 
sempre com uma velocidade determinada, independente do estado de movimento da 
fonte luminosa''.

Na melhor das hip\'oteses, em 1905 estes dois postulados poderiam ter sido 
considerados como alternativas poss\'\i veis. Hoje, depois de tantos testes 
experimentais e te\'oricos, n\~ao mais. 

Mas, na MR se insiste na eletrodin\^amica de Weber, por 
exemplo~\cite{mr6} %mr5 agora mr6

\begin{quotation}
As propriedades e vantagens da teoria eletromagn\'etica de Weber foram 
consideradas em outro livro.
\end{quotation} 

Essa teoria n\~ao tem nenhuma vantagem, ela j\'a foi descartada como 
proposta cient\'\i fica. Enfatizamos, um s\'eculo de experimentos e 
aplica\c c\~oes tecnol\'ogicas e, n\~ao menos importante, os esquemas 
conceituais constru\'\i dos a partir da eletrodin\^amica de Maxwell n\~ao 
deixam espa\c{c}o para ela. 
Lembre-se disso, caro leitor, quando assistir televis\~ao ou ouvir
a sua m\'usica favorita no seu {\em CD player}.

Da eletrodin\^amica se passa \`a gravita\c c\~ao, o autor continua
\begin{quotation}
... em analogia a eletrodin\^amica de Weber, propomos como a base para a 
mec\^anica relacional que a lei de Newton da gravita\c c\~ao universal seja 
modificada para ficar nos moldes da lei de Weber.
\end{quotation}

O leitor deve se convencer por ele mesmo que teorias de campo n\~ao
re\-la\-ti\-v\'\i s\-ti\-cas n\~ao est\~ao de acordo com a experi\^encia. 
O deslocamento Lamb e o momento magn\'etico do el\'etron e do m\'uon s\~ao 
exemplos, entre outros, da validade dessas 
teorias.~\footnote{Estes s\~ao os c\'alculos
esot\'ericos de teoria qu\^antica de campos mencionada por Will acima.} 

Mais ainda, os testes mais fortes, repetimos, de uma teoria s\~ao os 
indiretos. Por exemplo, toda a f\'\i sica de aceleradores n\~ao seria 
poss\'\i vel sem a TRE. Como foi mencionado na se\c c\~ao anterior, 
mesmo o sistema GPS est\'a usando ambas TRE e TRG.
Como explicar esse sucesso no contexto da MR?
 
O fato que os testes indiretos passam a ser mais importantes que os diretos 
(que s\~ao importantes quando se est\'a propondo uma teoria) faz com que 
caso algu\'em hoje repetisse
as experi\^encias de Michelson-Morley ou Fizeau e afirmasse ter achado
resultados opostos aos das experi\^encias originais, o experimento
ser\'a encarado como errado! [De fato isso aconteceu com a 
experi\^encia de Michelson-Morley: em 1926 um f\'\i sico chegou a 
conclus\~{o}es opostas. Nunca se confirmou onde estava o erro mas 
j\'a n\~ao era mais necess\'ario ach\'a-lo!]. \'E isso que quer dizer 
o conhecido f\'\i sico, premio Nobel de 1977, P. W. Anderson 
quando afirma que~\cite{pwa1} 
\begin{quotation}
It is the nature of physics that its generalizations are 
continually tested for correctness and consistency not only by careful 
experiments aimed directly at them but {\bf usually much more 
severely}~\footnote{Os negritos s\~ao nossos.}, by 
the total consistency of the entire structure of physics [$\cdots$]
My moral finally, is that physics--in fact all of science-- is a
pretty seamless web.
\end{quotation} 

A MR \'e uma teoria de tempo absoluto e n\~ao passa por testes que 
evidenciam a ``dilata\c c\~ao do tempo''. E mais, os argumentos (fracos) 
contra a TRE e TGR parecem ser motivados pelo fato do autor perceber que
quem as aceita n\~ao pode aceitar a MR. 

A dilata\c c\~ao do tempo em campos gravitat\'orios \'e particularmente
importante para demonstrar que os resultados da MR s\~ao inconsistentes
com as observa\c c\~oes. Seja por exemplo o caso de um corpo preso a uma 
mola oscilando horizontalmente. Este caso \'e considerado no 
MR~\cite{mr13} e o resultado \'e que a freq\"encia de oscila\c c\~ao 
\'e dada por
\begin{equation}
\omega=\sqrt{\frac{k}{m_g}}
\label{u2}
\end{equation} 
onde $k$ \'e a constante el\'astica da mola. Se observa no MR
que a diferen\c ca com o resultado na mec\^anica newtoniana \'e que na MR
aparece $m_g$ e n\~ao $m_i$.  O problema \'e quando
se usa o resultado da Eq.~(\ref{u2}) para afirmar~\cite{mr13}:
\begin{quotation}
Dobrando a quantidade de gal\'axias do universo, mantendo inalteradas
a mola, a Terra e o corpo de prova, diminuiria a freq\"u\^encia de 
oscila\c c\~ao em $\sqrt2$. Isto \'e equivalente a dobrar a massa inercial
newtoniana do corpo de prova.
\end{quotation}

Isto \'e, se extrapola um resultado que na pr\'atica coincide com o da
mec\^anica de Newton (\'e por isso n\~ao \'e importante) para o Universo 
todo! Qualquer sistema peri\'odico \'e um rel\'ogio. 
Acontece que se usamos isso para calcular a diferen\c ca de tempos de 
2 sistemas de molas, um na base e outro no alto de uma torre, a diferen\c ca
de tempos segundo a MR \'e: zero! Segundo a MR teriamos
\begin{equation}
\frac{\tau_2-\tau_1}{\tau_1}=\frac{N/\omega_2-N/\omega_1}{N/\omega_1}\equiv0
\label{ultima}
\end{equation}
onde $N$ \'e o n\'umero de oscila\c c\~oes e $\omega_{1,2}$ s\~ao a
freq\"uencias na base e no alto da torre. A identidade decorre da igualdade
entre $\omega_1$ e $\omega_2$ uma vez que pela Eq.~(\ref{u2}) 
as freq\"u\^encias so dependem da massa gravitacional do corpo a qual 
\'e inalter\'avel.~\footnote{Os autores agradecem G. E. A. Matsas discus\~oes
sobre este ponto. De fato, a Eq.~(\ref{ultima}) foi colocada pela 
primeira vez no debate realizado no IFGW-UNICAMP entre o autor da MR 
e Matsas. }
No entanto como mencionado acima
essa diferen\c ca dos rel\'ogios em campos gravitat\'orios j\'a foi bem
testada e est\'a em acordo com as teorias da relatividade. De fato, 
experimentos que medem o desvio para o vermelho gravitacional 
usando rel\'ogios em torres e o sistema GPS, como comentado acima, 
confirmam a TRG~\cite{gravitation2}.  

\'E interessante a afirma\c c\~ao com rela\c c\~ao \`a dilata\c c\~ao do 
tempo necess\'aria para explicar a chegada de p\'\i ons e m\'uons 
produzidos na atmosfera at\'e a Terra~\cite{mr7}:% (mr,157) 
\begin{quotation}
o mesmo pode ser aplicado na 
experi\^encia dos m\'esons. Ao inv\'es de afirmar que o tempo anda mais 
lentamente para o corpo em movimento, nos parece mais simples e de acordo 
com a experi\^encia afirmar que a meia-vida do m\'eson depende ou dos 
campos eletromagn\'eticos a que foi exposto nesta situa\c c\~ao ou ao 
seu movimento (velocidade ou acelera\c c\~ao) em rela\c c\~ao ao 
laborat\'orio e aos corpos distantes.
\end{quotation}
Acontece que a dilata\c c\~ao do tempo foi medida em circunst\^ancias
diversas: em aceleradores, em experimentos em avi\~oes e sat\'elites, 
em experi\^encias que medem o fator $g-2$ do m\'uon, etc. Quais campos 
eletromagn\'aticos se aplicariam nestes casos? Mesmo na atmosfera, se
existissem campos eletromagn\'eticos teriam outros efeitos por exemplo
nas comunica\c c\~oes via sat\'elite. 

Vemos ent\~ao que \'e a MR que n\~ao d\'a conta dos fatos observados 
como demostrado acima, um rel\'ogio na base de uma torre 
atrasa com rela\c c\~ao a um rel\'ogio no topo da mesma devido \`a 
influ\^encia do campo gravitacional da Terra. A  
Eq.~(\ref{ultima}) mostra ent\~ao que segundo a mec\^anica relacional, 
molas oscilando horizontalmente a Terra n\~ao
t\^em sua freq\"u\^encia, dadas pela Eq.~(\ref{u2}), 
alterada estejam elas na base ou no topo da torre. 
Tais osciladores seriam 
apenas um exemplo de rel\'ogios que n\~ao atrasariam devido ao 
campo gravitacional! A express\~ao dada na MR para a oscila\c c\~ao
depende apenas da massa gravitacional e da constante
da mola, que s\~ao conceitos primitivos em sua teoria 
e portanto n\~ao sofrem altera\c c\~ao das estrelas fixas.
	
Finalmente, a MR n\~ao \'e uma teoria de campos e portanto
n\~ao prev\^e a emiss\~ao de ondas gravitacionais de maneira natural. 
Apesar dos grandes detectores terrestres de ondas gravitacionais ainda 
n\~ao estarem em funcionamento, ondas gravitacionais j\'a foram 
indiretamente observadas em sistemas astrof\'\i sicos bin\'arios, 
como mencionado antes.

\section{Coment\'arios finais}
\label{sec:con}

O fato de uma teoria satisfazer ou n\~ao o Princ\'\i pio de Mach
(em qualquer uma de suas formula\c c\~oes) n\~ao pesa a favor ou contra
a teoria, visto que n\~ao h\'a qualquer experimento comprovando a validade
dele, mesmo porque seria bastante 
dif\'\i cil mover todas as estrelas do firmamento! 
A experi\^encia do balde n\~ao
pode ser considerada uma verifica\c c\~ao experimental do princ\'\i pio
como \'e afirmado na MR. 

Sabemos que a lei de Coulomb tem corre\c c\~oes de origem qu\^antica,
mas nem por isso dizemos que a lei deve ser mudada, apenas reconhece-se
que num determinado contexto (o \'atomo de hidrog\^enio, por exemplo)
outros fatores s\~ao importantes. No caso da lei da gravita\c c\~ao de 
Newton, que tinha sido testada para dist\^ancias maiores que 1 cm, 
pensava-se que poderiam ocorrer desvios para dist\^ancias da ordem de $\mu$m. 
Especula-se por exemplo, que, se existissem 
das dimens\~oes espaciais extras, o potencial gravitacional de Newton 
seria substitu\'\i do por uma express\~ao  mais geral. 
Trata-se de uma proposta te\'orica, por\'em medidas recentes na escala 
de 200 $\mu$m n\~ao mostram desvios da lei de gravita\c c\~ao de 
Newton~\cite{hoyle}.
Mesmo que esse tipo de teorias venha a 
ser confirmada no futuro ainda assim continuaremos a usar o potencial de 
Newton em muitas das aplica\c c\~oes em dist\^ancias de $\mu$metros at\'e 
milhares de quil\^ometros ou, dependendo da precis\~ao, a TRG. 
Mesmo que desvios da lei da gravita\c c\~ao fossem um dia observados, cabe 
ressaltar que seriam oriundos em teorias consistentes na maioria dos 
aspectos com a TRG. 
Estas teorias t\^em de acrescentar outros ingredientes te\'oricos, 
como simetrias extras ou mais dimens\~oes espaciais.
Por outro lado, na eletrodin\^amica qu\^antica temos o deslocamento Lamb, 
efeito bem medido no at\^omo de hidrog\^enio, que implica numa 
corre\c c\~ao ao potencial de Coulomb~\cite{halzen}.
O que aprendemos com estes exemplos? A resposta \'e que temos de ter sempre 
em mente em que contexto uma modifica\c c\~ao \'e feita numa lei b\'asica. 

Um aspecto que deve ser notado \'e que no pref\'acio da MR~\cite{mr8} 
aparece o seguinte %[mr, p. xviii]
\begin{quotation}
Este livro \'e direcionado a f\'\i sicos, matem\'aticos, engenheiros, 
fil\'osofos e historiadores da ci\^encia $[\cdots]$ Acima de tudo, 
\'e escrito para as pessoas jovens e sem preconceitos que t\^em interesse 
nas quest\~oes da f\'\i sica.
\end{quotation}
Deveria ser acrescentado {\it e com pouco senso 
cr\'\i tico}, porque para aceitar a MR, depois das observa\c c\~oes 
acima discutidas, \'e preciso, isto sim, ter preconceito {\it a favor} 
da mec\^anica relacional. Ainda no pref\'acio podemos ler~\cite{mr9}%  
%p. xvix
\begin{quotation} 
Ap\'os compreender a mec\^anica relacional entraremos num novo mundo,
enxergando os mesmos fen\^omenos com olhos diferentes e sob uma nova 
perspectiva. \'E uma mudan\c ca de paradigma.
\end{quotation}
O autor se refere ao conceito de paradigma cient\'\i fico introduzido 
po T. Kuhn. Sinceramente, leitor, se voc\^e
tem interesse nas quest\~oes da f\'\i sica, por acaso leu
em algum lugar 
que Einstein, Heisenberg, Bohr, Dirac, Fermi, Pauli, e tantos outros 
conhecidos cientistas fizeram logo de in\'\i cio esse tipo de 
afirma\c c\~ao? Vamos al\'em: esses autores escreveram livros sobre as suas
teorias somente depois de alguns anos e quando a comunidade de 
f\'\i sicos era  majoritariamente a favor delas.
Enfim, {\bf a verdadeira f\'\i sica nova} s\'o se percebe depois de certo 
tempo, mesmo para aqueles que a propuseram. Para Pais~\cite{ap6}
\begin{quotation} 
A nova din\^amica contida nas equa\c c\~oes relativistas 
ge\-ne\-ra\-li\-za\-das 
n\~ao foi completamente dominada, nem durante a vida de Einstein, nem no 
quarto de s\'eculo que se seguiu \`a sua morte [$\cdots$] nem mesmo num 
n\'\i vel puramente cl\'assico, ningu\'em pode hoje em dia gabar-se de 
ter um dom\'\i nio completo do rico conte\'udo din\^amico da din\^amica 
n\~ao linear designada por relatividade geral.
\end{quotation}

Apenas na {\it  ci\^encia patol\'ogica } as coisas s\~ao enganosamente 
claras de uma vez por todas. Ali\'as essa \'e, de fato, uma maneira de 
identific\'a-la. Os cientistas t\^em preconceitos, mas
mesmo estes est\~ao, na maioria dos casos, bem fundamentados.
Agora sabemos, por exemplo, que o esquema de Copernico-Kepler-Galileu 
precisava de uma f\'\i sica nova, afinal formulada por Newton; que 
o que Boltzmann queria n\~ao era poss\'\i vel sem a mec\^anica 
qu\^antica, que ainda n\~ao tinha sido 
descoberta.~\footnote{Mas seus m\'etodos e princ\'\i pios estavam 
corretos. Apenas a natureza n\~ao os realizava da maneira que ele
acreditava.}

Assim, n\~ao \'e apenas pelos testes diretos, desde Michelson e Morley, que 
a relatividade \'e aceita como correta em certo dom\'\i nio de fen\^omenos.
Mais importante ainda \'e a consist\^encia que ela trouxe para diversos
dom\'\i nios: astronomia, o sistema GPS, aceleradores de part\'\i culas,
f\'\i sica sub-at\^omica, etc.

Como exemplo de ci\^encia patol\'ogica
podemos lembrar do caso Velikovsky \cite{ilpf,eb}. Immanuel Velikovsky
(1895-1879) prop\^os uma teoria astron\^omica em seu livro ``Worlds in 
Collision''. Ali ele dava argumentos sobre uma s\'erie de cat\'astrofes
ocorridas na Terra, uma delas teria provocado a abertura do Mar Vermelho
para que os judeus vindos de Egito pudessem atravessar o mar. 
Ao ser questionado sobre a inexist\^encia de outros registros al\'em da 
B\'\i blia sobre esse tipo de cat\'astrofe
ele argumentava: ``amn\'esia coletiva provocada pelas mesmas cat\'astrofes''!
S\~ao esses argumentos de natureza {\em ad hoc}
que caraterizam a {\sl ci\^encia patol\'ogica}.
Em particular, segundo o qu\'\i mico e pr\^emio Nobel 
I. Langmuir~\cite{stone},   
a ci\^encia patol\'ogica tem a seguinte carater\'\i stica 
(existem outras mas estas s\~ao mais relacionadas com a ci\^encia 
experimental):  
\begin{quotation}
Fantastic theories contrary to experiences and criticisms are met by ad hoc 
excuses thought up on the spur moment.  
\end{quotation}
\begin{quotation}
Theorias fant\'asticas contr\'arias \`a experi\^encia e as cr\'\i ticas s\~ao
rebatidas com excusas ad hoc apropriadas para o momento
\end{quotation}

As cr\'\i ticas no livro MR \`as TRE e TRG s\~ao desse tipo, o argumento de 
``campos magn\'eticos'' para explicar a dilata\c c\~ao do tempo no caso 
dos raios c\'osmicos \'e caracter\'\i stico desse tipo de argumenta\c c\~ao 
{\em ad hoc}. J\'a foi discutido
na se\c c\~ao anterior que isso n\~ao procede, porque a dilata\c c\~ao
do tempo j\'a foi medida em diversas situa\c c\~oes e concorda bem com
a TRE e a TRG. Ao obs\-ti\-na\-da\-men\-te negar estas teorias e tudo que 
elas implicam, o autor da MR \'e obrigado a recorrer a processos 
misteriosos e invocar fontes ainda n\~ao investigadas, mas que tenham, 
para o incauto leitor,uma aura de plausibilidade 
(efeitos novos, campos magn\'eticos desconhecidos, etc.).
Por exemplo, no problema da precess\~ao das \'orbitas dos planetas para 
fixar o valor observado \'e introduzido de forma {\it ad hoc} um 
par\^ametro extra, $\xi$~\cite{mr11} %(mr, p.280) 
que deve valer $\xi=6$ para concordar com as medi\c c\~oes.

Por \'ultimo e n\~ao menos importante, o que dizer com rela\c c\~ao ao ensino
de f\'\i sica no terceiro grau? Como apresentar para os estudantes, sob 
o ponto de vista da MR, afirma\c c\~oes como as que seguem (tomadas do
respeitado e muito usado livro texto 
de Purcell~\cite{purcell2}) 
\begin{quotation}
Today we see in the postulate of relativity and
their implications a wide framework, one that embraces all physical laws and
not solely those of electromagnetism. We expect any complete physical theory
to be relativistically invariant.
\end{quotation}
\begin{quotation}
Hoje vemos nos postulados da relatividadee e suas implica\c c\~oes um 
referencial amplo, um que engloba todas as leis da f\'\i sica e n\~ao apenas as
da eletromagnetismo. Esperamos que qualquer teoria seja invariante relativista.
\end{quotation}

Ensinar-se-ia a um grupo de estudantes ``sem preconceitos'' que estas 
frases est\~ao erradas?, que toda a f\'\i sica do Sec.~XX tamb\'em est\'a?
N\~ao seria um crime deixar os estudantes nessa ignor\^ancia?

Mas n\~ao apenas no ensino em n\'\i vel de segundo e terceiro grau.
Por exemplo, n\~ao \'e conceb\'\i vel que um bi\'ologo, qu\'\i mico ou
f\'\i sico de outra especialidade, digamos de estado s\'olido ou 
ci\^encia dos materiais, use ferramentas como luz s\'\i ncroton e acredite
que a eletrodin\^amica de Weber ainda poderia ser considerada uma
teoria rival \`aquela que permitiu a constru\c c\~ao do aparelho que
usa nas suas pesquisas. 

Finalmente, gostar\'\i amos de observar o seguinte. Mesmo se nos 
restringirmos \`a {\it ci\^encia normal}~\footnote{Aqui usamos esse termo 
para rotular uma atividade de  pesquisa que se publica em revistas com 
razo\'avel par\^ametro de impacto.} podemos distinguir, numa mesma \'area,
diferentes comunidades. A primeira divis\~ao \'e pela especializa\c c\~ao.
Em geral uma comunidade tem uma ou v\'arias
revistas nas quais publica assuntos de um interesse que serve
para definir essa comunidade. A maioria das refer\^encias usadas no livro MR
est\~{a}o em revistas onde n\~ao s\~ao usualmente encontrados trabalhos
da {\it ci\^encia normal}. Se algu\'em tem argumentos v\'alidos de que
as TRE e TRG est\~ao erradas (esse ali\'as j\'a seria um resultado 
impressionante) deveria publicar em revistas como Physical
Review Letters. De nada a\-di\-an\-ta argumentar que essas revistas n\~ao 
publicariam, que t\^em preconceito etc.  
Isso mostra que as pessoas que apoiam os pontos de vista da MR pertencem a 
uma comunidade marginal. 

Para terminar, esperamos ter deixado claro duas coisas: 
{\it 1)} n\~ao \'e apenas pelos 
testes diretos, desde Michelson e Morley, que 
as teorias da relatividade s\~ao aceitas como corretas em certo 
dom\'\i nio de fen\^omenos.
Mais importante ainda \'e a consist\^encia que ela trouxe para diversos
dom\'\i nios: astronomia, aceleradores de part\'\i culas,
f\'\i sica sub-at\^omica, o sistema GPS etc.
{\it 2)} Que a proposta da mec\^anica relacional~\cite{mr} \'e errada e 
a resenha anterior~\cite{soares} \'e, por isso, inconseq\"uente.

\begin{center}
\vskip 1cm
{\bf Agradecimentos}
\end{center}

Agradecemos ao CNPq pelo auxilio financeiro parcial;  a L. F. dos Santos
pela leitura do manuscrito e a G. E. A. Matsas por \'uteis discus\~oes
sobre as teorias da relatividade. 

%\vskip .5cm

\end{document}